\input harvmac.tex
\noblackbox

\def\wdg{{ {\wedge} }}
\def\half{{ {1\over 2} }}

\def\tP{{ {\tilde P} }}
\def\tt{{ {\tilde t} }}

\def\tq{{ { \tilde q} }}

\def\frac#1#2{{#1\over #2}}
\def\one{{ {\bf 1} }}
\def\R{{ {\cal R} }}

\def\Q{{ {\cal Q} }}

\def\one{{ { {\bf 1}} }}
\def\tPhi{{ {\tilde \Phi} }}
\def\tF{{ {\tilde F} }}
\def\To{{ \rightarrow}}
\lref\OSV{H. Ooguri, A. Strominger, C. Vafa,
``Black Hole Attractors and the Topological String,''
Phys.Rev. D70 (2004) 106007, hep-th/0405146.}
 \lref\AOSV{M. Aganagic, H. Ooguri, N. Saulina, C. Vafa,
``Black holes, q-deformed 2d Yang-Mills, and non-perturbative
topological strings'',  Nucl. Phys. B715 (2005) 304-348, hep-th/0411280.}
\lref\AJS{M. Aganagic, D.
Jafferis, N. Saulina, `` Branes, black holes and topological strings
on toric Calabi-Yau manifolds'', JHEP 0612 (2006) 018, hep-th/0512245.}
\lref\AOO{M.
Aganagic, H. Ooguri, T. Okuda, `` Quantum Entanglement of Baby
Universes'', hep-th/0612067.} \lref\DGOV{R. Dijkgraaf, R. Gopakumar,
H. Ooguri, C. Vafa,`` Baby universes in string theory'', Phys.Rev. D73 (2006) 066002,
hep-th/hep-th/0504221.}
\lref\MD{A. Dabholkar, F. Denef, G. Moore,
B. Pioline, ``Exact and Asymptotic Degeneracies of Small Black
Holes'', hep-th/0502157, JHEP 0508 (2005) 021}
 \lref\DF{I. Bena, E. Diaconescu, B. Florea,
 ``Black string entropy and Fourier-Mukai transform'',hep-th/0610068.}
\lref\MDP{A. Dabholkar, F. Denef, G. Moore, B. Pioline, ``Precision
Counting of Small Black Holes'', hep-th/0507014, JHEP 0510 (2005)
096} \lref\Dab{A. Dabholkar, ``Exact counting of black hole
microstates'', hep-th/0409148.} \lref\Sen{A. Sen, ``Black holes,
elementary strings and holomorphic anomaly'', hep-th/0502126;
``Black-holes and the spectrum of half-BPS states in N=4
supersymmetric string theory'', hep-th/0504005.} \lref\Yin{D. Shih,
X. Yin, ``Exact Black Hole Degeneracies and the Topological
String'', hep-th/0508174.} \lref\ANV{M. Aganagic, A. Neitzke, C.
Vafa, ``BPS Microstates and the Open Topological String Wave
Function'', hep-th/0504054.}
 \lref\Denef{F. Denef, G. Moore,
 ``Split states, entropy enigmas, holes and halos'',
hep-th/0702146.}
 \lref\DM{E.Diaconescu,
G.Moore,``Crossing the wall: Branes vs. Bundles'', hep-th/0706.3193}
 \lref\Inaba{M.-A. Inaba, ``On the moduli of stable sheaves on a reducible
projective scheme and examples on a reducible quadric surface'',
Nagoya Math. J. 166 (2002), 135.}
\lref\Witrev{E. Witten, ``Two-dimensional gauge theories revisited,''
J. Geom. Phys. 9:303-368,1992, e-Print: hep-th/9204083.}

\Title{
  \vbox{\baselineskip12pt \hbox{arXiv:0710.0648}
  \hbox{}
  \vskip-.5in}
}{\vbox{
  \centerline{Fragmenting D4 branes and coupled q-deformed
  Yang Mills.
 }
\centerline{ }
  \centerline{}
}} \centerline{Daniel Jafferis$^1$, Natalia Saulina$^2$ }
\bigskip\medskip
\centerline{$^1${\it Department of Physics, Harvard University,
Cambridge, MA 02138}} \centerline{ {\it Department of Physics,
Rutgers University, Piscataway, NJ 08855} }
\bigskip
\centerline{$^2${\it Department of Physics, California Institute
of Technology, Pasadena, CA 91125}}
\medskip
\medskip
\medskip
\medskip
\medskip
We compute the index of BPS states for two stacks of D4-branes
wrapped on ample divisors and overlapping over a compact Riemann
surface of genus $g\ge 2$ inside non-compact Calabi-Yau 3-fold.
This index is given in terms of $U(N)\times U(M)$ q-deformed
Yang-Mills theory with bifundamental matter.
 From
the factorization in the limit of large D4 charge, we argue that
our result computes the jump in the index of BPS states across the
wall of the marginal stability for the split flow of a D4 brane
fragmenting into a pair of D4 branes.
\Date{October, 2007}
\newsec{Intoduction}
The conjecture \OSV\ of Ooguri, Strominger, Vafa relates
two important enumerative problems in mathematics and physics, the
counting of holomorphic curves in Calabi-Yau 3-folds and counting
of BPS degeneracies of four dimensional black holes. In
formulating the OSV conjecture for non-compact manifolds, one
replaces BPS states of black holes with BPS states of D-branes
wrapped on various cycles inside the manifold.

In a non-compact setup, OSV conjecture was successfully tested for
$D4$ branes wrapped on an ample divisor $O(-p)\rightarrow
\Sigma_g$ inside $O(-p)+O(p+2g-2)\rightarrow \Sigma_g$ \AOSV\ and
for systems of D4-branes wrapped on ample divisors inside toric
Calabi-Yau's and intersecting over non-compact Riemann surface
\AJS. More recent work \AOO\ involves both $D4$-branes wrapped on
an ample divisor and anti-$D4$ branes wrapped on a non-ample
divisor and develops the connection with baby universes proposed
in \DGOV.

In the compact setup, the conjecture for black holes preserving
four supercharges was tested to leading order in \MD\MDP\Yin\DF .
The conjecture was found to have extensions to half BPS black
holes in compactifications with ${\cal N}=4$ supersymmetry
\Dab\Sen\MD\MDP. In \ANV\ the version of the conjecture for open
topological strings was formulated.

The original motivation for our work was to test OSV conjecture
for the system of $N$ $D4$ branes wrapped on ample divisor $D_1$
and $M$ $D4$ branes wrapped on ample divisor $D_2$ intersecting
over compact Riemann surface. At first we were discouraged to find
that the index of BPS degeneracies $Z$ does not have a large
charge limit consistent with OSV. Instead, in this limit we
obtained schematically
$$Z \sim \vert Z^{top}(t_N)\vert^2  \vert Z^{top}(t_M)\vert^2+\ldots  $$
where $Z^{top}$ is topological string partition sum
and $t_N$($t_M$) indicates that
Kahler modulus of Riemann surface is fixed to attractor value
for $N D4$ branes wrapped on $D_1$($M D4$ branes wrapped on
$D_2.$) For the complete expression for $Z$ in this limit see eq.(4.3).

However, inspired by recent results for split attractor flows
\Denef\DM, we realized that our computation should be interpreted
as giving the jump in the index of BPS states across the wall of
marginal stability for the split $D4\rightarrow D4+D4.$ The
Vafa-Witten theory with bifundamentals on the intersecting D4
branes determines the contribution to the partition function of a
D4 wrapping the total class that arises from the corner of moduli
space where the D4 has fragmented into these two pieces. The
computation of marginal stability line for our system is similar
to \DM\ but we include it in Section 5 to show that it can be
crossed by starting from  large values of background K\"ahler
modulus where both unfragmented and fragmented $D4'$s contribute
to the BPS index.

This note is organized as follows. In Section 2 we derive the
contribution of the fragmented $D4$ brane to the index of BPS
degeneracies in terms of $U(N)\times U(M)$ q-deformed Yang-Mills
with bifundamental matter. In section 3 we take large D4 charge
limit and do saddle point analysis to identify configurations
giving the dominant contribution to the index. In Section 4 we
expand the index around this dominant contribution and observe
factorization appropriate for $D4\rightarrow D4+D4$ split. Section
5 gives evidence for  interpretation of our result as the jump in the index
of BPS states across the wall of marginal stability.

\newsec{Bound states of D4-branes intersecting over Riemann surface.}
Let us consider a non-compact Calabi-Yau three-fold
$O(-p+g-1)\oplus O(p+g-1)\rightarrow \Sigma_g,$ that can be
thought of as the local neighborhood the intersection curve of two
surfaces in a compact manifold. For $p=0$ this local model
describes behavior near the canonical divisor $\Sigma_g$ of the
complex surface ${\cal P}$ inside $O(-K)\rightarrow {\cal P}.$ We
further assume $g>1$ and $0\le p < g-1$ so that both
 divisors $D_1=O(-p+g-1)\rightarrow \Sigma_g$
and $D_2=O(p+g-1)\rightarrow \Sigma_g$ have deformations.

Wrap N D4 branes on $D_1$ and M D4 branes on $D_2$ so that these
two stacks of branes overlap over $\Sigma_g.$ We now compute the
partition function $Z_g$ which counts (with sign) bound states of
these branes. The D0 and D2 brane charges induced by bifundamental
matter and fluxes on the branes are weighted by chemical
potentials.

It is natural to expect that in this local geometry, the coupled
four dimensional Vafa-Witten theory localizes to a 2D theory on
$\Sigma_g$. This reduced theory must be q-deformed Yang Mills with
gauge group $U(N) \times U(M)$ coupled to bifundamental matter.
The path integral has the form \eqn\pathint{Z_g={1\over N!M!}\int
d\Phi \, d\tPhi \, dX_{bf} \Delta_h^{2-2g}(\Phi) \,
\Delta_h^{2-2g}(\tPhi) \, e^{-S^{(N)} -S^{(M)} -S_{bifund}} }
where the q-YM measure was derived in \AOSV\
$$ \Delta_h(\Phi)=\prod_{1\le i < j \le N}
\bigl(e^{i{(\Phi_i-\Phi_j)\over 2}}-e^{i{(\Phi_j-\Phi_i)\over 2}}\bigr),\quad
\Delta_h(\tPhi)=\prod_{1\le a < b \le M}
\bigl(e^{i{(\tPhi_a-\tPhi_b)\over 2}}-e^{i{(\tPhi_b-\tPhi_a)\over 2}}\bigr)
$$
and the q-YM action is
\eqn\defin{S^{(N)}={\theta \over g_s} \int_{\Sigma_g}
 Tr\Phi\wdg \omega+
{1\over g_s} \int_{\Sigma_g} Tr F \Phi +{g-1-p\over 2g_s}
\int_{\Sigma_g} \omega \wdg Tr\Phi^2}
\eqn\definii{S^{(M)}={\theta \over g_s} \int_{\Sigma_g}
 Tr\tPhi\wdg \omega+
{1\over g_s} \int_{\Sigma_g} Tr \tF \tPhi +{g-1+p\over 2g_s}
\int_{\Sigma_g} \omega \wdg Tr\tPhi^2}
In \pathint\ $dX_{bf}$ denotes the measure for
bi-fundamentals and in \defin,\definii\ $\omega$ stands for the unit-volume K\"ahler form on $\Sigma_g$.

We now specify the action for bifundamentals $S_{bifund}$ and
integrate them out in the path-integral.
Our system of intersecting branes has four ND directions
(these are directions normal to $\Sigma_g$ in $CY_3.$)
Hence from flat space analysis we know that there are
bifundamental  fermion and  bifundamental scalar.
The fermion is a section of the bundle
$$S_{\Sigma}\otimes E_N \otimes E_M^*\oplus S_{\Sigma}\otimes E_N^* \otimes E_M$$
and boson is a section of the bundle
$$S_+\otimes E_N \otimes E_M^* \oplus S_-\otimes E_N^* \otimes E_M$$
Here $E_N$ denotes $U(N)$ vector bundle over Riemann surface
$\Sigma_g,$ while $S_{\Sigma}$ is spin bundle over it. Finally,
$S_{\pm}$ is chiral (anti-chiral) spin bundle associated with the
normal bundle to $\Sigma_g$ in $CY_3$. After performing a
topological twist we find the topological bifundamentals:
$(M_1,\mu_1)$ are 0-forms with values in $E_N \otimes E_M^*,$
$(h_1,v_1)$ are $(0,1)$ forms with values in $E_N \otimes E_M^*,$
$(M_2,\mu_2)$ are 0-forms with values in $E_N^* \otimes E_M,$
$(h_2,v_2)$ are $(0,1)$ forms with values in $E_N^* \otimes E_M.$
For $k=1,2$ $M_k$ and $h_k$ are bosons, $\mu_k$ and $v_k$ are
fermions. The BRST transformations of bifundamental fields are
\Witrev\ :
$$ \delta_Q M_k=\mu_k, \quad \delta_Q \mu_k=-i\bigl(\Phi\otimes \one-
\one \otimes {\tilde \Phi}\bigr)M_k,\quad k=1,2$$
$$ \delta_Q v_k=h_k, \quad \delta_Q h_k=-i\bigl(\Phi\otimes \one-
\one \otimes {\tilde \Phi}\bigr)v_k,\quad k=1,2.$$
Recall also BRST transformations of the basic multiplet of cohomological
2D Yang Mills theory \Witrev\ :
$$\delta_QA_{\mu}=i\psi_{\mu},\quad \delta_Q \psi_{\mu}=-D_{\mu}\Phi,\quad
\delta_Q\Phi=0$$
 $$\delta_Q{\tilde A}_{\mu}=i{\tilde \psi}_{\mu},\quad \delta_Q {\tilde \psi}_{\mu}=-D_{\mu}\tPhi,\quad
\delta_Q{\tilde \Phi}=0$$
We consider the Q-exact action for bifundamentals(we skip writing appropriate
trace in the formulae below):
$$S_{bifund}=S_{bifund}^{(1)}+S_{bifund}^{(2)},\quad
S_{bifund}^{(k)}=\{ Q, V_k+tV_k'\} \quad k=1,2$$
$$V_k=
\int_{\Sigma_g} \Bigl({\bar v}_k D_{\bar z} M_k + D_{z}({\bar
M}_k) v_k\Bigr),\quad V_k'=t \int_{\Sigma}\Bigl( e {\bar
\mu}_kM_k+{\bar v}_k h_k\Bigr),\quad k=1,2.$$ where $e={\sqrt g}.$
Since $S_{bifund}$ is Q-exact the partition function is
independent of the parameter $t$, hence we are free to take it
large to render the bifundamental fields heavy. Below we integrate
out bifundamentals $M_1,\mu_1,v_1,h_1$ and at the end comment on
the result for integrating out fields $M_2,\mu_2,v_2,h_2.$

The action $S^{(1)}_{bifund}$ has the
form:
$$ S^{(1)}_{bifund}=\int_{\Sigma} \Bigl( t {\bar h}_1 h_1+{\bar h}_1D_{\bar z}\,M_1+
D_{z}{\bar M}_1 h_1+t e {\bar \mu}_1\mu_1 +
t e {\bar M}_1\bigl(\Phi\otimes \one-\one \otimes {\tilde \Phi}\bigr)M_1$$
$$+t {\bar v}_1 \bigl(\Phi\otimes \one-
\one \otimes {\tilde \Phi}\bigr)v_1-
{\bar v}_1\bigl(\psi_{\bar z}-{\tilde \psi}_{\bar z}\bigr)M_1+
{\bar M}_1\bigl(\psi_z-{\tilde \psi}_z\bigr)v_1-{\bar v}_1 D_{\bar z}\mu_1 +
D_{z} {\bar \mu}_1 v_1 \Bigl)$$
Integrating out $h_1$ and changing variable
$$w_1=v_1-{1\over t}\bigl(\Phi\otimes \one-
\one \otimes {\tilde \Phi}\bigr)^{-1}\Bigl(D_{\bar z} \mu_1+\bigl(\psi_{\bar z}-{\tilde \psi}_{\bar z}\bigr)M_1\Bigr)$$
we find the remaining action
$$ S^{(1)}_{bifund}=t \int_{\Sigma}\Biggl[  e \Bigl( {\bar \mu}_1\mu_1 +
{\bar M}_1\bigl(\Phi\otimes \one-\one \otimes {\tilde \Phi}\bigr)M_1 \Bigr)
+{\bar w}_1\bigl(\Phi\otimes \one-\one \otimes {\tilde \Phi}\bigr)w_1 \Biggr] $$
$$ -{1 \over t}\int_{\Sigma}  \Biggl[ D_{z}{\bar M}_1 D_{\bar z} M_1 +
\Bigl(D_{z}{\bar \mu}_1+{\bar M}_1\bigl(\psi_z-{\tilde \psi}_z\bigr)\Bigr)\bigl(\Phi\otimes \one-
\one \otimes {\tilde \Phi}\bigr)^{-1} \Bigl(D_{\bar z} \mu_1+\bigl(\psi_{\bar z}-{\tilde \psi}_{\bar z}\bigr)M_1\Bigr)\Biggr]$$
In the limit of large $t$ we can drop the terms of order $O({1\over t})$ in the action.
Let us introduce a basis of eigenmodes of $D_z D_{\bar z}$:
$$ D_z D_{\bar z} f^I=\lambda_I \epsilon_{z{\bar z}} f^I$$
Then we expand
$$M_1=\sum_I M_{1(I)}f^I,\quad \mu_1=\sum_I \mu_{1(I)} f^I.$$
For non-zero eigenvalue $\lambda_I\ne 0$ we also define
$g_{\bar z}^I={1\over \sqrt{\lambda_I}} D_{\bar z} f^I$ and expand  $w_1$
as
$$w_{1{\bar z}}=\sum_{I: \lambda_I\ne 0}w_{1(I)} g_{\bar z}^I + \sum_{a=1}^{h^1} w_{1a} g_{\bar z}^a$$
Here $h^m=dim H^m(\Sigma, E_N \otimes E_M^*)$ and $g_{\bar z}^a$ for $a=1,\ldots h^1$ are zero modes.
Altogether, we find that integrating out bifundamentals $M_1,\mu_1,v_1,h_1$
contributes to
the measure for $\Phi$ and ${\tilde \Phi}$:
$$ det^{h^1-h^0} \bigl(\Phi\otimes \one-\one \otimes {\tilde \Phi}\bigr)$$

Analogously, integrating out bifundamentals $M_2,\mu_2,v_2,h_2$
contributes to
the measure for $\Phi$ and ${\tilde \Phi}$:
$$ det^{k^1-k^0} \bigl(\Phi\otimes \one-\one \otimes {\tilde \Phi}\bigr)$$
where we denote $k^m=dim H^m(\Sigma_g, E_N^* \otimes E_M).$
Now we use Riemann-Roch theorem to compute
$$h^1-h^0=g-1+c_1(E_M)-c_1(E_N),\quad k^1-k^0=g-1+c_1(E_N)-c_1(E_M)$$
where $c_1(E_N)$ is the first Chern class of bundle $E_N.$
We conclude that integrating out all bifundamentals contributes
to the measure \eqn\bif{\Sigma^{2(g-1)}(\Phi,\tPhi)} where
$$\Sigma(\Phi,\tPhi)=
 \prod_{i=1}^N
\prod_{b=1}^M \bigl(e^{i{(\Phi_i-\tPhi_b)\over 2}}-
e^{i{(\tPhi_b-\Phi_i)\over 2}}\bigr)$$ and we took into account
that  $\Phi,{\tilde \Phi}$ are periodic.

Hence the total measure in the resulting path-integral over
$\Phi,{\tilde \Phi}$ has the form
\eqn\meas{{\cal G}(\Phi,\tPhi)={\Delta_h^{2-2g}(\Phi)
\Delta_h^{2-2g}(\tPhi) \over \Sigma^{2(1-g)}(\Phi,\tPhi)}} and the
path integral \pathint\ is brought to the form
\eqn\pathint{Z_g={1\over N!M!}\int d\Phi \, d\tPhi \, {\cal
G}(\Phi,\tPhi)\, e^{-S^{(N)} -S^{(M)}} } After summing over all
flux configurations
$$F_i=2\pi r_i \omega, \quad \tF_a=2\pi s_a \omega,\quad r_i,s_a \in {\bf Z}\quad i=1,\ldots,N,\quad a=1,\ldots,M$$
path integral localizes to
$$\Phi_i=ig_sn_i,\quad {\tilde \Phi}_{b}=ig_sm_{b}$$
and we find
\eqn\finres{Z_g={1\over N!M!}\sum_{n_i, m_a \in {\bf Z}}
{\cal G}\bigl(ig_s{\vec n},ig_s{\vec m}\bigr)
 q^{-{g-1-p \over 2} {\vec n}^2 -
{g-1+p\over 2} {\vec m}^2} e^{i \theta \bigl( \sum_{i=1}^N n_i+\sum_{a=1}^M m_a\bigr) }}
where $q=e^{-{g_s}}.$
Let us rewrite partition function $Z_g$  as a sum over
representations $\R$ of $U(N)$ and $\Q$ of $U(M).$ \eqn\ours{
Z^{(N,M)}_g=\Upsilon \sum_{\R-U(N),\, \Q-U(M)}\, (Z_{\R
\Q})^{2-2g} \tq^{{g-1-p\over 2} C_2(\R)+ {g-1+p\over 2} C_2(\Q)}
e^{i \theta \bigl(C_1(\R)+C_1(\Q)\bigr)} } where $\tq=q^{-1}$ is
small expansion parameter and \eqn\crq{Z_{\R
\Q}=\tq^{NC_1(\Q)+MC_1(\R)\over 2}\sum_{A-SU(M)} S^{(M)}_{A
\Q}(\tq)\, S^{(N)}_{A \R}(\tq) } and sum goes over Young diagrams
$A$ with number of rows less than M . We have assumed $N\ge M$ and
used
$$\Sigma^{-1}\bigl(ig_s{\vec n},ig_s{\vec m}\bigr)=
\tq^{N C_1(\Q)+M C_1(\R)\over 2}\,
\sum_{A-SU(M)}\, Tr_A(\tq^{\Q+\rho^{(M)}})\,Tr_A(\tq^{\R+\rho^{(N)}})$$
as well as the definition of S-matrix
$$S_{A \Q}(q)=Tr_{\Q}(q^{\rho^{(M)}})\,Tr_A(q^{\Q+\rho^{(M)}})$$
where $\rho^{(N)}_i={N-2i+1\over 2}$ is the Weyl vector of $U(N).$
The overall normalization, $\Upsilon$, is ambiguous and can be
fixed from the requirement of factorization of $Z_g$ in the large
N,M limit similar to \AOSV.

\subsec{Cap in the holonomy basis} One can think of the partition
function \ours\ as obtained from an operatorial approach, i.e. by
sewing $2g-2$ pants. Then the cap in the holonomy basis is given
by \eqn\holbas{C(U,V)=\sum_{\R,\, \Q} Tr_{\R}U Tr_{\Q}V Z_{\R
\Q}=}
$${1\over \Delta_h(u) \Delta_h(v)}\sum_{\mu} \sum_{w \in S_M} \sum_{\sigma \in S_N}
(-)^{w+\sigma} \delta\Bigl(v+i{g_s N\over 2}
+ig_sw(\mu+\rho^{(M)})\Bigr) \,\, \delta\Bigl(u+i{g_s M\over 2}
+ig_s\sigma(\mu+\rho^{(N)})\Bigr) $$ where $U=e^{iu},\quad
V=e^{iv}.$

Recall that, as discussed in \AJS, insertion of the operator
$Tr_{\R} e^{i\Phi}$ into the path-integral of the $U(N)$ q-YM on
the cap gives \eqn\insr{Z_N(C,Tr_{\R}e^{i\Phi})={1\over
\Delta_h(u)}\sum_{\sigma \in S_N} (-)^{\sigma} \, \delta\Bigl(u
+ig_s \sigma(\R+\rho^{(N)}) \Bigr)} So that the cap $C(U,V)$ is
simply
$$C(U,V)=\sum_{\mu - SU(M)}
Z_N(C,Tr_{\mu}e^{i\Phi})\,\, Z_M(C,Tr_{\mu}e^{i\tPhi})$$ We can
motivate this by consistency as in \AJS, assuming for simplicity
that $N=M=1$.

The operator insertion \insr\ enforces the delta function in
\holbas\  which, in the case $N=M=1$, simply says that $$u=-v.$$
Recall that the flux of the Vafa-Witten gauge field living on each
of the D4 branes is computed in the two dimensional reduction by
the holonomy of the gauge field around the boundary of the cap,
$$u = \int_{Cap} F^{(1)} = -\int_{Cap} F^{(2)} = - v.$$ Therefore,
along the intersection cap, the fluxes must be the same.

It is suggestive that in the mathematical description of the
classical theory, namely the moduli space of sheaves with support
on the intersecting pair of divisors, a similar condition arises.
In particular, stable coherent sheaves supported on the union of
$D_1$ and $D_2$ are defined by a sheaf on $D_1$ and one on $D_2$,
together with an isomorphism from ${\cal E}_1|_{\Sigma_g} \To
{\cal E}_2|_{\Sigma_g}$, as shown in section 3 of \Inaba. In the
rank 1 case, this isomorphism serves to identify the fluxes along
$\Sigma_g$, precisely as we find in the cap amplitude. The two
dimensional theory we have constructed should thus be regarded as
the correct quantization of the classical description of this
moduli space of sheaves.

Note that the quantization in units of $g_s$ that is also encoded
by \holbas\ is automatic in the q-deformed Yang-Mills theory, thus
aside from the positivity of the row lengths of $\mu$, the
insertion \insr\ is implied by the natural $U(N) \times U(M)$
covariant generalization of the requirement of equal flux along
the intersection curve.

\newsec{Saddle point analysis.}
Suppose that the genus, $g \geq 2$, and that $\theta=0$. Then we
will determine the dominant contribution to the partition function
\finres\ in the large $N$ and $M$ limit.

Let us first set $p=0.$
The term $\frac{1}{\Delta_h (n)^{2g-2}}$ results in an attractive
force between the $N$ eigenvalues, which are also pushed to the
bottom of the quadratic potential, $g_s (g-1) n^2 /2$. The same
holds for the $M$ eigenvalues. The effect of the opposite
statistics of the bifundamental multiplets is that the $n_i$ repel
the $m_j$. Recalling that all the $N+M$ eigenvalues must be
distinct, we see that the dominant contribution consists of a
clump of $N$ and a clump of $M$, either touching or separated by
some distance. These two possibilities correspond, respectively,
to the existence of two or four Fermi seas.

Let us parameterize the position of the clumps by their midpoints
at position $x$ for the $N$ and $y$ for the $M$. Clearly, the
attractive interaction among the $n_i$ is unaffected by variation
of $x$, and likewise for the $M$ eigenvalues. The repulsive force
is given by $(g-1) g_s \coth \left( \frac{g_s}{2} (n_i - m_j)
\right) $ between two eigenvalues of opposite type. This is
strictly stronger than the constant force of $g_s (g-1)$, which is
approached in the limit $n_i - m_j >> 0$.

We will use this approximation $[n-m]_q \sim e^{|g_s (n-m)|/2}$,
hence the actual separation is at least at great as what we find
here. The action of the clump of $N$ depends on $x$ in the large
$N$ limit as $$\frac{g_s (g-1)}{2} \sum_{i=-N/2}^{N/2} (x+i)^2
\approx \frac{g_s (g-1)}{6} \left( (x+\frac{N}{2})^3 -
(x-\frac{N}{2})^3 \right) = g_s (g-1) x^2 N / 2, $$ up to terms
independent of $x$ or subleading in $1/N$. Putting this together
with the constant repulsive force between the clumps (which gives
a total of $N M g_s (g-1)$ ), we find the saddle point conditions
$$g_s (g-1) N x = g_s (g-1) M N $$ $$g_s(g-1) M y = - g_s (g-1) M
N.$$ This has solution $x = M$ and $-y = N$, which has two clumps
whose centers are separated by distance $N+M$.

For $0< p < g-1$ we find analogously:
$$x_p={(g-1)M\over g-1-p},\quad y_p=-{(g-1)N\over g-1+p}$$

Please note that working in the regime $q^{-1} \ll 1$ is crucial for
the existence of the saddle point. There is no saddle point in the
other regime $q \ll 1.$

\newsec{Large $N$ and $M$ factorization for $g \geq 2$.}
To take the large $N$ and $M$ limit of the partition sum $Z_g$ we
use \finres.
 Given
the picture we just determined of the dominant contribution to the
partition function \finres\ , it is natural to
 parametrize
the eigenvalues as
$${\vec n}=x_p+\rho^{(N)}+R_+{\overline R_-}[l_R],\quad {\vec m}=
y_p-\rho^{(M)}-Q_+{\overline Q_-}[l_Q]$$
 where  $U(N),U(M)$ representations are written
 in terms of coupled representations
$$\R=R_+{\overline R_-}[l_R],\quad \Q=Q_+{\overline Q_-}[l_Q]$$
For simplicity we assume that $N$ and $M$ are even.

To make contact with OSV conjecture, we do analytic continuation
and consider $q$ as a small expansion parameter below. Then we
have
$$\eqalign{\frac{\Sigma (ig_s {\vec n}, ig_s {\vec m})}
{\Delta_h (ig_s {\vec n}) \Delta_h (ig_s {\vec m})} =
\frac{q^{-{NM(x_p-y_p)\over 2}} q^{-{N M\over 2} (l_R+l_Q)}
q^{-{N\over 2} (|Q_+|-|Q_-|)}q^{-{M\over 2}(|R_+|-|R_-|)} }{ S_{0
[R_+{\overline R_-}]} \ S_{0 [Q_+ {\overline Q_-}]} }\cr \prod_{i
= 1}^{N/2} \prod_{j=1}^{M/2} (1-q^{x_p-y_p+l_R+l_Q} q^{(\rho_N^+ +
R_+)_i + (\rho_M^+ +Q_+)_j} ) (1-q^{x_p-y_p+l_R+l_Q} q^{(\rho_N^+
+ R_+)_i - (\rho_M^+ +Q_-)_j} ) \cr (1-q^{x_p-y_p+l_R+l_Q}
q^{-(\rho_N^+ + R_-)_i + (\rho_M^+ +Q_+)_j} )
(1-q^{x_p-y_p+l_R+l_Q} q^{-(\rho_N^+ + R_-)_i - (\rho_M^+ +Q_-)_j}
),}$$ where $\rho_N^+ = (\frac{N-1}{2}, \frac{N-3}{2}, \dots,
\frac{1}{2} )$ is the positive half of the Weyl vector, $\rho_N$.

Recalling the Schur function identity, $$\sum_\eta s_\eta (x)
s_{\eta^t} (y) = \prod_{i,j} (1+x_i y_j),$$ we see that each of
the interaction terms between the four Fermi surfaces can be
expanded in the form $$ \prod_{i,j} (1-q^{x_p-y_p+l_R+l_Q}
q^{(\rho_N^+ + R_+)_i + (\rho_M^+ +Q_+)_j} ) = \sum_\eta s_\eta
(q^{\rho_N^+ + R_+}) (-)^{|\eta|} q^{(x_p-y_p+l_R+l_Q)|\eta|}
s_{\eta^t} (q^{\rho_M^+ +Q_+}).$$ It is crucial that $\eta$ is a
Young diagram, not a $U(N)$ representation, hence it has positive
rows lengths, and must be small due to the suppression by
$q^{(x_p-y_p)|\eta|}$. This implies that $s_\eta (q^{\rho_N^+ +
R_+}) \rightarrow q^{N |\eta|/2} \frac{W_{\eta
R_+}(q)}{W_{R_+}(q)}$ in the large $N,M$ limit. For convenience,
let the distance between the clumps be denoted by $d_p = x_p - y_p
= \frac{g-1}{(g-1)^2 - p^2} \bigl( (g-1)(N+M) - p(N-M) \bigr)$.

Therefore, using the factorization formula for
$S_{0 [R_+{\overline R_-}]}$ derived in \AOSV,
we find that in the large $N$ and $M$ limit:
$$\eqalign{ \Sigma (ig_s {\vec n}, ig_s {\vec m})=
q^{- N M d_p / 2} q^{-\frac{N M}{2} (\ell_R + \ell_Q)}
q^{-\frac{N}{2}( |Q_+|-|Q_-|)}q^{-\frac{M}{2}(|R_+|-|R_-|)} \cr
\sum_{\eta_{a b}} (-)^{\sum |\eta_{a b}|} e^{-\sum_{a,b} \bigl(
(t_a-g_s \ell_R) + (\tt_b-g_s \ell_Q) \bigr) |\eta_{a b}|}
\prod_{a,b} \frac{W_{\eta_{a b} R_a}(q^a)}{W_{R_a}(q^a)}
\frac{W_{\eta_{a b}^t Q_b}(q^b)}{W_{Q_b}(q^b)}  },$$ where $a,b =
\pm$ parametrize the couplings between pairs of Fermi seas, the
W-functions are evaluated at either $q$ or $q^{-1}$ as indicated,
and we have defined $$\eqalign{t_a = g_s \frac{(g-1) N}{g-1+p} +
(-)^a g_s \frac{N}{2} \cr \tt_b = g_s \frac{(g-1) M}{g-1-p} +
(-)^b g_s \frac{M}{2}. }$$

The other piece in $Z_g$  in the large N,M limit
has the form
$$q^{-\half \Bigl(x_p^2(g-1-p)+y_p^2(g-1+p)\Bigr)}
\, q^{-M(g-1)\bigl(Nl_R + \vert R_+\vert -\vert R_-\vert\bigr)} \,
q^{-N(g-1)\bigl(Ml_Q + \vert Q_+\vert -\vert Q_-\vert\bigr)}$$
$$\times q^{-{(g-1-p)\over 2}C_2\Bigl(R_+{\overline R_-}[l_R]\Bigr)}
\, q^{-{(g-1+p)\over 2}C_2\Bigl(Q_+{\overline Q_-}[l_Q]\Bigr)} $$
Using expressions for quadratic Casimirs of $R_+{\overline
R_-}[l_R]$ and $Q_+{\overline Q_-}[l_Q]$ (see for example \AOSV)
we find \eqn\ourfact{ Z_g\sim \vert Z_{top}(t) \vert ^2 \, \vert
Z_{top}(\tt) \vert ^2 +\ldots} where the Kahler modulus in the
first (second) $Z_{top}$ factor is attractor value for $N$ D4
branes wrapped on $D_1$ ($M$ D4 branes wrapped on $D_2$).
\eqn\ourk{t={(g-1+p)N \over 2}+i\theta_1,\quad \tt={(g-1-p)M \over
2}+i\theta_2}

Fixing the normalization factor by requiring the partition
function to factorize, we need $\Upsilon = \alpha(g_s, \theta; N)
\alpha(g_s, \theta; M) q^{N M (g-1) d_p /2}$, where $\alpha$ is
defined in \AOSV; the classical pieces of the prepotential we find
will be identical to the ${\hat Z}_0$ of \AOSV. Putting everything
together, we have the full factorization formula:
$$\eqalign{Z = \sum_{P_i \tP_i \eta_{a b}^i} Z_N^+(g_s; t- (g-1-p) g_s \ell_R, t^+ - g_s
\ell_R) \ Z_N^-(g_s; \bar{t} + (g-1-p) g_s \ell_R, t^- - g_s
\ell_R) \cr Z_M^+(g_s; \tt-(g-1+p) g_s \ell_Q, \tt^+ - g_s \ell_Q)
\ Z_M^-(g_s; \bar{\tt} + (g-1+p) g_s \ell_Q, \tt^- - g_s
\ell_Q),}$$ where the chiral blocks are defined by
$$Z^+_N (g_s; t, t^+) = {\hat Z_0}(t) e^{-\frac{t}{g-1+p} \sum |P_i|} e^{-t^+ \sum |\eta_{+ b}^i|} \sum_R
{ q^{\frac{1}{2}(g-1+p) \kappa_R} e^{-t|R|} \over W_R^{2g-2}}
\prod_{i=1}^{2g-2} \bigl( \frac{W_{P_i R}}{W_R}
\frac{W_{\eta_{++}^i R}}{W_R} \frac{W_{\eta_{+-}^i R}}{W_R}
\bigr)$$
$$Z^-_N (g_s; t, t^-) = {\hat Z_0}(t) e^{-\frac{t}{g-1+p} \sum |P_i|} e^{-t^- \sum |\eta_{- b}^i|}  \sum_R
{ q^{\frac{1}{2}(g-1+p) \kappa_R} e^{-t|R|} \over W_R^{2g-2}}
\prod_{i=1}^{2g-2} \bigl( \frac{W_{P_i R}}{W_R}
\frac{W_{{\eta_{-+}^i}^t R^t}}{W_{R^t}} \frac{W_{{\eta_{--}^i}^t
R^t}}{W_{R^t}} \bigr)$$
$$Z^+_M (g_s; \tt, \tt^+) = {\hat Z_0}(\tt) e^{-\frac{\tt}{g-1-p} \sum |\tP_i|}
e^{-\tt^+ \sum |\eta_{a +}^i|} \sum_R { q^{\frac{1}{2}(g-1-p)
\kappa_R} e^{-\tt|R|} \over W_R^{2g-2}} \prod_{i=1}^{2g-2} \bigl(
\frac{W_{\tP_i R}}{W_R} \frac{W_{{\eta_{++}^i}^t R}}{W_R}
\frac{W_{{\eta_{+-}^i}^t R}}{W_R} \bigr)$$
$$Z^-_M (g_s; \tt, \tt^-) = {\hat Z_0}(\tt) e^{-\frac{\tt}{g-1-p} \sum |\tP_i|}
e^{-\tt^- \sum |\eta_{a -}^i|} \sum_R { q^{\frac{1}{2}(g-1-p)
\kappa_R} e^{-\tt|R|} \over W_R^{2g-2}} \prod_{i=1}^{2g-2} \bigl(
\frac{W_{\tP_i R}}{W_R} \frac{W_{{\eta_{+-}^i} R^t}}{W_{R^t}}
\frac{W_{{\eta_{--}^i} R^t}}{W_{R^t}} \bigr).$$ This can be
expressed in a more suggestive form by trading the sums over
``ghost'' representations for integrals over $SU(\infty)$ matrixes
associated to the noncompact moduli. In particular, we have:
\eqn\factor{\eqalign{Z = \sum_{\ell_R, \ell_Q} \int d {\vec U} d
{\vec V} d {\vec W_{\pm \pm}} \psi(g_s; t-(g-1-p) g_s \ell_R,
{\vec U}, {\vec W_{+ \pm}}) \bar\psi(g_s; t+(g-1-p) g_s \ell_R,
{\vec U}, {\vec W_{- \pm}}) \cr \psi(g_s; \tt-(g-1+p) g_s \ell_Q,
{\vec V}, {\vec W_{\pm +}}) \bar\psi(g_s; \tt+(g-1+p) g_s \ell_Q,
{\vec V}, {\vec W_{ \pm -}}),}} where the contour integrals are
over matrices of the form $U = e^u$ for $u$ with fixed real part
given by the attractor value of the noncompact moduli: $$Re [u] =
\frac{2 \ t}{g-1+p}, \ \ Re[v] =\frac{2 \ \tt}{g-1-p}, \ \ Re[w_{a
b}] = t^a + \tt^b.$$

\newsec{Fragmenting 4-branes}
We conjecture that $U(N)\times U(M)$ q-deformed Yang-Mills theory
with bifundamental matter is the microscopic worldvolume
description of those states in the BPS Hilbert space of a single
D4 brane that correspond to a split attractor flow in
supergravity. Moreover, these are precisely the contributions to
the Euler character of the D4-brane moduli space which come from
the corner where the brane wrapping ample divisor $D=ND_1+MD_2$
fragments into $N$ $D4$ branes on $D_1$ and $M$ $D4$ branes on
$D_2$.

Consider a 4-brane wrapping a very amply decomposable divisor, $D
= D_1 + D_2 \in H^2(X)$, in the Calabi-Yau. Lifting to M-theory,
the partition function with chemical potentials turned on for the
D2 and D0 fluxes is given by the MSW CFT that is essentially a
$(0,4)$ sigma model into the moduli space of deformations of the
4-brane. The exact formulation of this conformal field theory is
unknown, but the dominant classical contribution is believed to
reduce to the Euler character of the nonsingular part of the
classical moduli space of a surface in the class $[D]$.

We are interested in understanding a particular subleading
correction to the partition function that comes from the singular
corner of the 4-brane moduli space where the surface splits into a
pair wrapping $D_1$ and $D_2$ and intersecting over the common
curve, $\Sigma_g$. In particular, we argue that this contribution
to the entropy is best analyzing using the Vafa-Witten worldvolume
theory living on the 4-branes, and the topological bifundamental
matter coupling them along the intersection. Furthermore, we
speculate that this piece of the full partition function is
associated to certain two centered black holes states in the
supergravity limit. Therefore our partition function should be
interpreted as the microscopic quantum theory describing the jump
in the index of BPS states across the wall of marginal stability
for the $D \To D_1 + D_2$ split, at least in the case of local
Calabi-Yau.

In compact models \DM\ demonstrated that the marginal stability
line for $D4\rightarrow D4+D4$ split can be crossed by starting a
flow from large values of background K\"ahler modulus
$t_{\infty}$. We now show that this is also the case
 in our non-compact model. Let $\Gamma_1$($\Gamma_2$) be the
charge of N D4-branes (M D4-branes) wrapped on $D_1$($D_2$). Let
$\omega_i$ be Poincare Dual of divisor $D_i$ for $i=1,2$. Then the
charges including fluxes on the branes are given by
$$\Gamma_1=N\omega_1+(\half N^2 \omega_1^2 -f_1
\omega_1\omega_2)+q_0 dV$$
$$\Gamma_2=M\omega_2+(\half M^2 \omega_2^2 -f_2
\omega_1\omega_2)+q_0' dV,$$ where $dV$ is the volume form, and we
use the same sign conventions as \DM.

Let $B+iJ=T_1\omega_1+T_2\omega_2$ be the complexified K\"ahler form.
In our geometry we know intersection numbers
$$C_{112}=g-1-p,\quad C_{122}=g-1+p$$
Furthermore, the numbers $C_{111}$ and $C_{222}$ (which are a
priori ambiguous due to non-compactness) can be extracted from
\AOSV\ where classical contribution to free energy was fixed, i.e.
we relate
$${T^3\over (g-1+p)(g-1-p)}=C_{111}T_1^3+C_{122}T_1T_2^2+C_{211}T_2T_1^2+C_{222}T_2^3$$
where $T=T_1C_{112}+T_2C_{122}$ is the complexified K\"ahler modulus
of Riemann surface $\Sigma_g.$
This procedure gives
$$C_{111}={C_{112}^2\over C_{122}},\quad C_{222}={C_{122}^2\over C_{112}}$$

Now we can the compute intersection pairing for the charges which
is generically non-zero for $f_1\ne f_2$. For example, if $p=0$,
then
$$\langle \Gamma_1, \Gamma_2 \rangle=
(g-1)(\half NM^2 -Nf_2-\half MN^2 +Mf_1)\ne 0.$$

Let us consider the region in the moduli space with $ImT_1\gg 1,\quad ImT_2\gg 1.$
Then, we evaluate central charges $Z_1=Z(\Gamma_1)$ and $Z_2=Z(\Gamma_2)$ as
$$Z_1=-\half {NT^2\over C_{122}}+T\bigl(\half {N^2 C_{112}\over C_{122}}-f_1\Bigr)
-q_0$$
$$Z_2=-\half {MT^2\over C_{112}}+T\bigl(\half {M^2 C_{122}\over C_{112}}-f_2\Bigr)
-q_0'$$ Please note that dependence on the complexified K\"ahler
moduli $T_1,T_2$ in both $Z_1$ and $Z_2$ comes only via the
complexified K\"ahler modulus $T$ of $\Sigma_g.$

Let us write $T=x+iy$ and compute $Im(Z_1 {\bar Z}_2)$:
$$Im(Z_1 {\bar Z}_2)=y\Bigl({\alpha \over 2} y^2 +\beta(x)\Bigr)$$
where
$$\alpha=\half {MN^2\over C_{122}}-\half {NM^2\over C_{112}}+
{Nf_2\over C_{122}}-{Mf_1\over C_{112}}$$
$$\beta(x)=\alpha x^2 +x \Bigl({Nq_0'\over C_{122}}-{Mq_0\over C_{112}}\Bigr)
+q_0\Bigl({M^2 C_{122}\over 2 C_{112}}-f_2\Bigr)-q_0'\Bigl({N^2 C_{112}\over 2 C_{122}}-f_1\Bigr)$$

To  find a solution $(x_{MS},y_{MS})$ of
$Im(Z_1 {\overline Z_2})=0,$ with $y_{MS}\gg 1$
we need to have
$${\beta(x)\over \alpha}<0.\quad \vert {\beta(x)\over \alpha} \vert \gg 1 $$
This can clearly be done for a choice of B-field and 2-form fluxes
$f_1,f_2$ as well as a choice of flux-induced D0 charges $q_0,q_0'.$
For example, consider $N=M$ and $p=0.$ Let us further assume $x^2\ll N,$
choose $f_2-f_1$ to be of order 1 and positive, $q_0-q'_0$ to be of order
1 and negative and $q'_0f_1-q_0f_2$ be of order much less than $N^2.$
With this choices, we find
$$\beta={N^2 (q_0-q_0')\over 2}<0$$
$$\alpha={N\over g-1}(f_2-f_1)>0$$
so that the ratio  ${\beta \over \alpha}$ is of order $N$ and negative.

The above consideration demonstrates the existence of marginal stability
wall for $D4\rightarrow D4+D4$ split only for large $N,M,$
however we find it plausible that such a wall exists even for finite $N,M.$
As an evidence that our results may be related with split flows of \Denef\
even for finite $N,M$ note the following. Let us consider $N=M=1$ and $p=0.$
Then, inserting the factor $\Sigma^{2g-2}$ in the measure of
path-integral in eq.(2.5)(the result of integrating out
bifundamentals) in the limit
of small coupling $g_s$ is equivalent to
\eqn\Splitii{g_s(g-1)\Biggl(Z^{(M)} \frac{\del}{\del \theta} Z^{(N)} - Z^{(N)}
\frac{\del}{\del \theta} Z^{(M)}\Biggr) .} But this is nothing
else but inserting angular momentum degeneracy
$$\langle \Gamma_1,
\Gamma_2 \rangle = (g-1) \left(N Q_2 - M Q_1 \right)$$
 as in \DM,
since the derivatives with respect to $\theta$ simply pull down
factors of $\frac{1}{g_s} Q_{D2}$. It would be interesting to
investigate this connection further.
Below we propose a tentative
interpretation of the factor $\Sigma^{2g-2}$(for $N=M=1$ and
$p=0$) as arising from the halo of $D2/D0$-branes surrounding the
two D4-fragments.

Consider $D2/D0$ system with charge and central
charge given by
$$\Gamma_3=-f_3 \omega_1\omega_2+q_3[pt],\quad Z_3=-f_3T-q_3.$$
Let us find solutions of $Im(Z_1 {\bar Z}_3)=0,$ where
$$Im(Z_1 {\bar Z}_3)=y\Bigl({\gamma \over 2} y^2 +\delta(x)\Bigr)$$
with
$$\gamma={Nf_3\over C_{122}},\quad \delta(x)=\gamma x^2 +
x \Bigl({Nq_3\over C_{122}}\Bigr) -q_0f_3-q_3\Bigl({N^2
C_{112}\over 2 C_{122}}-f_1\Bigr)$$ Let us look for solutions in
the region $x^2 \ll N.$ We may choose $f_3$ to be positive of
order 1, $q_3f_1-q_0f_3 \ll N^2$ and $q_3>0.$ Then, the ration
${\delta \over \gamma}$ is negative and of order $N.$ Hence, there
exists marginal stability wall for the charge $\Gamma_1$ and
$\Gamma_3$ precisely in the same region-- $x^2\ll N$ and $y^2\sim
N$ --as MW for the $\Gamma_1,$ $\Gamma_2$ charges found above.
Again, this argument holds for large $N,M$ but let us assume
that the conclusion is also true for finite $N,M,$ i.e.
there exists a halo of $D2/D0$ particles, which are mutually BPS with the two fragmented
D4-branes. If each such particle has one unit of D2 charge
wrapping $\Sigma_g$ and $n-m$ units of D0 charge, then the angular
momentum degeneracy is $\langle \Gamma_1+\Gamma_2,\Gamma_3
\rangle =2g-2.$ If each such particle is a fermion and u is
chemical potential, then one finds the factor
$(1-u^{n-m})^{2g-2}.$ Our factor $\Sigma^{2g-2}$ is a modular
transformed version of this.

There is a subtlety in the generating function for the index of
split states in a noncompact geometry, which modifies \Splitii. As
explained by \ANV, the entropy computed by the q-deformed
Yang-Mills is in the mixed ensemble for the local D0 and D2
charges, but the D2 charges associated to 2-cycles that become
noncompact in the local limit are held fixed. Formally, one has
the relation $$Z_{local}(\phi_0, \phi; F) = \int du
Z_{mixed}(\phi_0, \phi, u) e^{F \cdot u},$$ where $F \in H_2(X)$
such that $F \cdot [\Sigma_g] = 0$ is the fixed noncompact D2
flux, and $u$ are the associated chemical potentials.

Therefore one expects that the local split partition function will
be determined in terms of the fragments via
$$\eqalign{Z_{local}(\phi_0, \phi; F) = \sum_{f \in H_2} \big( N
Z_{local}^1 (\phi_0, \phi; f) D_1 \cdot \left( F - f- \del \right)
Z_{local}^2 (\phi_0, \phi; f-F)  \cr - M Z_{local}^2 (\phi_0,
\phi; f) D_2 \cdot \left( f - \del \right) Z_{local}^1 (\phi_0,
\phi; f) \big),}$$ where the sum is over $f \cdot [\Sigma_g] = 0$
that can be supported as flux on $D_1$ and $F-f$ can be obtained
as flux on $D_2$.

In the large charge limit, this is exactly the form that we find
in the factorization of the coupled q-deformed Yang-Mills \factor,
where the ``ghost representations'' parametrize these noncompact
flux sectors. The integrals over the noncompact moduli serve to
enforce the condition that the noncompact D2 charge bound as flux
in the two fragments cancel.

\vskip 1cm {\bf Acknowledgments}

We are grateful to D.E. Diaconescu, D. Gaiotto, G. Moore, H.
Ooguri, and C. Vafa for useful discussions. It is our pleasure to
thank the Stony Brook physics department and the 4th Simons
Workshop in Mathematics and Physics for hospitality during some
stages of this work. Research of D.J. was supported in part by NSF
grants PHY-0244821 and DMS-0244464. Research of N.S. was supported
in part by DOE grant DE-FG03-92-ER40701.

\listrefs
 \bye